\documentclass[a4paper,usenatbib,twocolumn]{mnras}
\usepackage{bm}
\usepackage{amsmath}

\usepackage{multirow}

\usepackage{color}
\usepackage{bm}
\usepackage{color}
\usepackage{mathrsfs}  
\usepackage{dcolumn}
\usepackage{graphicx}
\usepackage{hyperref}

\newif\ifAMStwofonts
\AMStwofontstrue

\def\vxi{{\bm \xi}}

\def\dk{\delta^{\mathrm{K}}}

\def\vnabla{\bm{\nabla}}

\def\gsim{~\rlap{$>$}{\lower 1.0ex\hbox{$\sim$}}}

\def\simpropto{\lower.2ex\hbox{$\; \buildrel \propto \over \sim \;$}}
\def\ltsim{\lower.5ex\hbox{$\; \buildrel < \over \sim \;$}}
\def\gtsim{\lower.5ex\hbox{$\; \buildrel > \over \sim \;$}}
\def\ltsim{\lower.5ex\hbox{$\; \buildrel < \over \sim \;$}}
\def\gtsim{\lower.5ex\hbox{$\; \buildrel > \over \sim \;$}}

\def\pa{\partial}

\def\vnabla{{\bf \nabla}}

\def\kms{\mbox{km\,s$^{-1}$}}

\def\dd{\,{\rm d}}

\def\kms{\ {\rm km\,s^{-1}}}
\def\hmpc{\ {\rm h^{-1}Mpc}}

\def\dd{{\rm d}}

\def\ln{{\rm ln}}
\def\pa{\partial}

\def\la{\langle}
\def\ra{\rangle}

\def\pmb#1{\setbox0=\hbox{#1}%
\kern-.025em\copy0\kern-\wd0
\kern.05em\copy0\kern-\wd0
\kern-.025em\raise.0433em\box0}

\def\vv{\pmb{$v$}}

\def\vv{\pmb{$v$}}

\def\vx{\pmb{$x$}}
\def\vy{\pmb{$y$}}

\def\vr{\pmb{$r$}}
\def\tvr{\tilde {\pmb{$r$}}}
\def\tr{\tilde {r}}

\def\tmu{\tilde \mu}

\def\hvr{\hat {\vr}}
\def\htvr{\hat {\tvr}}

\def\hvx{\hat {\vx}}
\def\hvy{\hat {\vy}}
\def\hy{\hat {y}}
\def\hr{\hat {r}}
\def\htr{\hat {{\tilde r}}}

\def\simlt{\lower.5ex\hbox{$\; \buildrel < \over \sim \;$}}
\def\simgt{\lower.5ex\hbox{$\; \buildrel > \over \sim \;$}}

\def\vnabla{\pmb{$\nabla$}}

\newcommand{\dsc}[2]{{#1}_{_{#2}}}
\newcommand{\beq}{\begin{equation}}
\newcommand{\eeq}{\end{equation}}
\def\beqa{\begin{eqnarray}}
\def\eeqa{\end{eqnarray}}
\def\fixit#1{}
\def\hmpc{h^{-1}\,{\rm Mpc}}

\def\dd{{\rm d}}

\def\physrep{Physics Reports}

\def\aap{A\&A}

\title[Velocity-Density correlations ]{Velocity-Density  Correlations from the \textit{cosmicflows-3}  Distance  Catalog
and the 2MASS Redshift Survey}
\author[A. Nusser]{Adi Nusser\thanks{\href{adi@physics.technion.ac.il}{adi@physics.technion.ac.il}}
\thanks{Physics Department and the Asher Space Science Institute-Technion, Haifa 32000, Israel}}

\date{\today}
\begin{document}
\label{firstpage}
\pagerange{\pageref{firstpage}--\pageref{lastpage}}

\maketitle
\begin{abstract}
The peculiar velocity of a mass tracer is on average aligned with the dipole modulation of the surrounding mass density  field. 
We present a first measurement of the correlation between  radial peculiar velocities of objects in the \textit{cosmicflows-3}  catalog
 and the dipole moment of the 2MRS galaxy distribution in concentric spherical shells centered on these  objects. 
Limiting the analysis to \textit{cosmicflows-3} objects with  distances of $100\hmpc$, the correlation function  is detected at  a confidence level  $\gtsim 4\sigma$.
The measurement is found  consistent with the  standard $\Lambda$CDM model  at   $\ltsim 1.7\sigma$ level. We formally derive the constraints
$0.32<\Omega^{0.55}\sigma_8<0.48$ ($68\% $ confidence level) or equivalently 
$0.34<\Omega^{0.55}/b<0.52$, where $b$ is the galaxy bias factor.  
Deeper and improved peculiar velocity catalogs will substantially reduce the uncertainties, allowing tighter  constraints  from this type of  correlations.
 \end{abstract}
\section{Introduction}

The peculiar velocity of a galaxy is expected to be correlated with its surrounding matter distribution. 
A classic example is the motion of the Local Group (LG) of galaxies relative to the frame of reference defined by the temperature anisotropy of the cosmic microwave background (CMB).
The dipole component of the CMB temperature implies a motion of $V_{\mathrm lg}=627\pm 22\kms$ in the direction $(l,b)=(276^\circ\pm 3^\circ,30^\circ \pm 3^\circ)$  \citep[e.g.][]{kogut1993}. In accordance with gravitational instability theory for structure formation \citep{peeb80}, this motion is produced 
by the cumulative  gravitational pull of  the observed large scale structure as probed by a variety of surveys of the galaxy distribution 
\citep[e.g][]{sd88,LL89,schmoldt99,chodor04,bilicki11,carrick15}. Any deviations can  easily be explained  within the limitations of current surveys \citep{Nusser2014}. 
More generally, the analysis of \cite{Davis2011} demonstrated a  tight correlation between the observed radial velocities of galaxies in the SFI++ \citep{spring07} catalog and the gravitational force field corresponding to the galaxy distribution in the $ K_s\le 11.25$ Two Mass Galaxy Redshift survey (2MRS) \citep{2mrs2012}.

The  analyses above required a derivation of the gravitational field from a given distribution of 
galaxies in a redshift survey. In the case of the 
radial peculiar velocity versus gravity comparison, equivalent smoothing of the two fields is also essential. 
Here we aim at a simpler approach which can be implemented more directly, alleviating the complex steps in the 
analysis above. We aim at a direct measurement of  the $u-\delta$  correlation between radial motion, $u$,  of galaxies  and 
the dipole moment of the density contrast, $\delta$,  on spherical shells centered on the galaxies. 
We will  perform the correlation using the radial peculiar velocities from the \textit{cosmicflows-3}  catalog \citep{cf3}
and the galaxy distribution in the  $K_s\le 11.75$ 2MRS. These two data sets match in depth and 
are the best available  in terms  of  sky coverage. 

This $u-\delta$  correlation greatly  mitigates  several complications inherently present in the more detailed velocity-gravity comparison. 
As we shall see, by taking the dipole moment of the correlation function, the  ``Kaiser rocket"  effect \citep{k88} resulting from the flux limit nature of the redshift survey is 
essentially eliminated.
The estimation of the correlation does not require any smoothing of either the peculiar velocity data or the galaxy distribution. 

The outline of the paper is as follows. In \S\ref{sec:basic} we provide definitions for the basic quantities. The theoretical framework 
and predictions in the $\Lambda$CDM model are presented in \S\ref{sec:theory}. The main properties of the observational data,  \textit{cosmicflows-3} and 2MRS, are briefly described in \S\ref{sec:obs}).
This section also gives details of how the correlations are estimated from the observations. Errors associated with the observed correlations will 
be assessed using  mock galaxy catalogs as described in \S\ref{sec:mocks}. The results of $u-\delta$ correlations and implications on cosmological parameters  follow in \S\ref{sec:results}.  In \S\ref{sec:vvdelta} we provide a preliminary assessment of 3D velocity density correlations.  We conclude with a general discussion in \S\ref{sec:conc}.

Our notation is as follows. The expansion parameter is $a$ and is related to the redshift  $ z$ by $a=(1+z)^{-1}$. The Hubble 
function  is $H(t)=\dot a/a $ and is denoted by $H_0$ at the present epoch.
We adopt the $\Lambda$CDM cosmological model denoting by 
$\Omega$ and $\Lambda$ the mass energy density parameters in units of the critical density.
The growth factor of linear density  perturbations is  $D(t)$ and the growth rate is 
$f (z) = \dd \ln D/\dd \ln a$. The growth rate mainly depends on the mass density 
parameter $\Omega(z)$ and is well approximated by $f\approx \Omega^{0.55}$  in a $\Lambda$CDM cosmology \citep{Lind05}. 
Further, we  define the correlation function $\dsc{\xi}{f g}\equiv \la f(\vr)g(\tvr)\ra $ between any two fields $f$ and $g$ 
at  points $\vr$ and $\tvr$ as the ensemble average over many random realizations of the 
fields. 

\section{Basics}
\label{sec:basic}

The density contrast is defined as $\delta(\vr)=\rho(\vr)/\bar \rho-1$, where $\bar \rho $ is the mean background density.  The  comoving coordinate of a patch of matter is $\vr$ and the corresponding comoving peculiar velocity is $\vv$ (physical is $a\vv$). 
Linear theory  of cosmological gravitational instability \citep{peeb80} relates $\vv$ and $\delta$ via
\begin{equation}
\label{eq:linr}
\delta(\vr)=-\frac{1}{f}\vnabla \cdot \vv(\vr)\; ,
\end{equation}
where distance coordinates are expressed in $\kms$.
This, however, is the real density contrast corresponding to mass per unit volume $\dd^3 r=r^2\dd r \dd \Omega$ in actual distance space.
Surveys of the galaxy distribution   provide redshifts, $cz$ (in $\kms$), of galaxies rather than distances. The latter are provided in distance indicator catalogs  for a smaller set of galaxies. 
The observed redshifts are given in terms of the comoving distance, $r$, and  the radial component $u=\vv \cdot \hvr $ of $\vv $ \citep{sw},
\begin{equation}
\label{eq:czsw}
cz=cz_c+u
\end{equation}
where $z_c$ is the cosmological redshift calculated from the actual comoving distance\footnote{If $\dsc{r}{\mathrm{Mpc}}$ is the comoving distance in Mpc then $dr=Hd\dsc{r}{\mathrm{Mpc}}=cdz_c$
},  $r$, of the galaxy and, since we are considering low redshift data only, 
we have neglected gravitational contributions.
By continuity and to linear order,  density contrast of matter in redshift space, $\delta^s$, 
differs from $\delta$ by the divergence of the displacement.  Therefore,
\begin{equation}
\label{eq:ds}
\delta^s=\delta-\vnabla \cdot (u \hvr) \; .
\end{equation}
We write 
\begin{equation}
\label{eq:divu}
\vnabla \cdot (u \hvr)=\hat r_i \hat r_j \frac{\partial v_i}{\partial x_j}+\frac{2 u}{r}\; ,
\end{equation}
where $\hat r_i $ is  cosine the angle between $ \hvr$ and the cartesian axis defined by the unit vector $\hvx_i$ and summation over repeated indices is implied.
The derivatives and all other quantities are taken as a function of the real space coordinate $r$ which, to linear order, are equivalent to 
quantities expressed in terms of coordinates in redshift space.

Therefore,  observational data of radial peculiar motions and the distribution of galaxies in redshift space 
allow a direct  estimation of the correlation function
\begin{equation}
\label{eq:xisud}
\dsc{\xi}{u\delta}^s=\la u(\vr) \delta^s(\tvr) \ra \; .
\end{equation}
rather than $\dsc{\xi}{u\delta}=\la u(\vr) \delta(\tvr) \ra$ between $u$  and the real space density field, $\delta$.
The superscript $s$  on $\xi$ will be reserved for velocity-density correlation  in redshift space.
 
The relation (\ref{eq:ds}) implies
\begin{equation}
\label{eq:xisud}
 \dsc{\xi}{u \delta}^s=\xi_{u \delta}+\dsc{\xi}{wu} -\frac{2 \xi_{uu}}{\tr}\; ,
\end{equation}
where 
$w=-\htr_i  \htr_j {\partial v_i}/{\partial \tilde x_j} $.
 We begin  with a theoretical description of $\dsc{\xi}{u\delta}^s$ in linear theory and then we proceed to compute these correlations from mock and real data. 
 
\section{Theoretical correlations}
 \label{sec:theory}
  Expressions for each of the terms on the r.h.s  of (\ref{eq:xisud}) can be conveniently derived  
using the formalism developed by  \cite{Gorski1988} for describing  velocity correlations.
The correlation of the  cartesian 
components $v_i$ and $v_j$ at different points in space is 
\begin{equation}
\label{eq:gorski}
\dsc{\xi}{v_i v_j}=\Psi_\perp(y)\delta^{\mathrm K}_{ij}+\left[\Psi_\parallel(y)-\Psi_\perp(y)  \right]\hat {y}_i \hat {y}_j\; , 
\end{equation}
where 
$ \delta^{\mathrm K}_{ij}$ is Kronecker's delta function.
The ``perpendicular"  and ``parallel" (to the separation $\vy=\tvr -\vr$)  correlation functions $\Psi_\perp$ and $\Psi_\parallel$  are given in terms of the density power spectrum, $P(k)$, as 
\begin{equation}
 \Psi_{\perp,\parallel}(y)=\frac{f^2}{2\pi^2}\int_0^\infty P(k) K_{\perp,\parallel}(k y)\dd k\; ,
\end{equation}
where $K_\perp(x)=j_1(x)/x$ and  $K_\parallel(x)=j_0(x)-2j_1(x)/x$ and $j_l(x)$  are  spherical Bessel functions of order $l$.

For  $\dsc{\xi}{uu}=\la \hr_i v_i(\vr) \htr_j v_j(\tvr)\ra $,  the relation (\ref{eq:gorski})
yields
\begin{equation}
\label{eq:uu}
\dsc{\xi}{uu}= \nu \Psi_\perp+\mu \tmu\Psi_-\; ,
\end{equation}
where $\Psi_-= \Psi_\parallel -\Psi_\perp$, $\mu=\hvr \cdot \hvy$,  $\tilde \mu=\htvr \cdot \hvy$ and $ \nu=\hvr\cdot \htvr$.
To derive $\dsc{\xi}{wu}$, we 
write \footnote{$\partial y/\partial \tilde x_i=\hy_i $ and 
$\pa \hy_i/\pa \tilde x_j=\dk_{ij}/y =\hy_i\hy_j/y$ where $y=\sqrt{\sum_i (\tilde x_i -x_i)^2}$.}
\begin{eqnarray}
\label{eq:indices}
\la\frac{\partial v_i}{\partial \tilde x_j}v_m \ra&=&\Psi'_\perp\hy_j \delta^{\mathrm{K}}_{im}+\Psi'_-\hy_i \hy_j \hy_m\\
\nonumber 
&+&\frac{\Psi_-}{y}\left(\delta^{\mathrm{K}}_{ij}\hy_m +\delta^{\mathrm{K}}_{jm}\hy_i\right)-2\frac{\Psi_-}{y}\hy_i \hy_j \hy_m\; .
\end{eqnarray}
where primes denote derivatives with respect to $y$. 
Therefore, contracting with $\hr_i \hr_j$ gives
\begin{equation}
\label{eq:Xi}
\dsc{\xi}{wu}=-\left[\Psi'_\perp (2\tilde \mu\; \nu +\mu)+(\Psi'_--2\Psi'_\perp){\tilde \mu}^2 \mu\right]\; .
\end{equation}
 
According to the  linear  relation (\ref{eq:linr}), the correlation of $u$ with real space density is   $\dsc{\xi}{u\delta}=-\dsc{\xi}{\theta u}/f$ where  $\theta =\vnabla \cdot \vv=\delta^{\mathrm{K}}_{ij}\frac {\partial v_i}{\partial \tilde x_j}$. From  (\ref{eq:gorski})  we have, 
\begin{equation}
\label{eq:thetav}
\dsc{\vxi}{\vv \theta}=\left(3\Psi'_\perp+\Psi'_- \right)\hvy\; ,
 \end{equation}
where we have used    $\Psi_-=y \Psi'_\perp$ \citep{Gorski1988}.
Thus, $\dsc{\xi}{u\theta}=\la  u(\vr)\theta(\tvr) \ra= \hvr \cdot \dsc{\vxi}{\vv \theta}$  and hence
\begin{equation}
\label{eq:thetau}
\dsc{\xi}{ u\delta} =-\frac{1}{f}\left(3\Psi'_\perp+\Psi'_- \right)\mu\; .
\end{equation}
For later use we also list
\begin{equation}
\dsc{\xi}{uv_j}=\la \hr_i v_i(\vr)v_j(\tvr)\ra =\Psi_\perp\hr_j +\mu\Psi_-\hy_j\; .
\end{equation}

 Collecting all terms, we arrive at
\begin{eqnarray}
\dsc{\xi}{u\delta}^s(\vr,\tvr)&=&-\frac{1}{f}\left(3\Psi'_\perp+\Psi'_- \right)\mu \\
\nonumber &-&3 \Psi'_\perp \mu 
 -\Psi'_\perp (2\tilde \mu\; \nu +\mu)-(\Psi'_--2\Psi'_\perp){\tilde \mu}^2 \mu
\\
 \nonumber &-&2\frac{ \nu \Psi_\perp+\mu \tmu\Psi_-}{\tr}\; ,
\end{eqnarray}
where $\nu$ and $\tmu$ are expressed in terms of $y$, $r$ and $\mu$.

The form of the dependence  on $\mu$  suggests a multipole expansion of 
the correlations by means of  Legendre polynomials, $P_l(\mu)$\footnote{$P_0=1$, $P_1=\mu$, $P_2=(3\mu^2-1)/2$ and $P_3=(5\mu^3-3\mu)/2$. },
Coefficients of the expansion can be computed easily in the 
separations $y/r\ll 1$, i.e. the distant observer limit (DOL),  where $\nu\approx 1$ and 
$\tilde \mu=\mu$. We will consider $\la \dsc{\xi}{u\delta}^s \ra_l=1/(4\pi)\int \dd \Omega \dsc{\xi}{u\delta}^sP_l(\mu)\dd \mu=1/2\int \dd \mu \dsc{\xi}{u\delta}^sP_l(\mu)$. Evaluating the integrals gives
\begin{eqnarray}
\la \dsc{\xi}{u\delta}^s \ra_0&=&-\frac{ 2\Psi_\perp+ 2\Psi_-/3}{\tr}\\
\la \dsc{\xi}{u\delta}^s \ra_1&=&-\left(\frac{3}{5}+\frac{1}{f}\right)\Psi'_\perp-\left(\frac{1}{5}-\frac{1}{3f}\right)\Psi'_-\\
\la \dsc{\xi}{u\delta}^s \ra_2&=&-\frac{4}{15}\frac{\Psi_-}{\tr}\\
\la \dsc{\xi}{u\delta}^s \ra_3&=&\frac{2}{35}\left(2\Psi'_\perp -\Psi'_-\right)\; ,
\end{eqnarray}
and all higher moments vanish identically. The term  $2\dsc{\xi}{uu}/\tr$ contributes only to 
$l=0$ and $l=2$ and does not  affect any of the other multipoles. Because of the $1/r$ dependence, this term 
mainly depends  on nearby data. 

\begin{figure}
\begin{centering}  
\includegraphics[angle=0,width=0.5\textwidth]{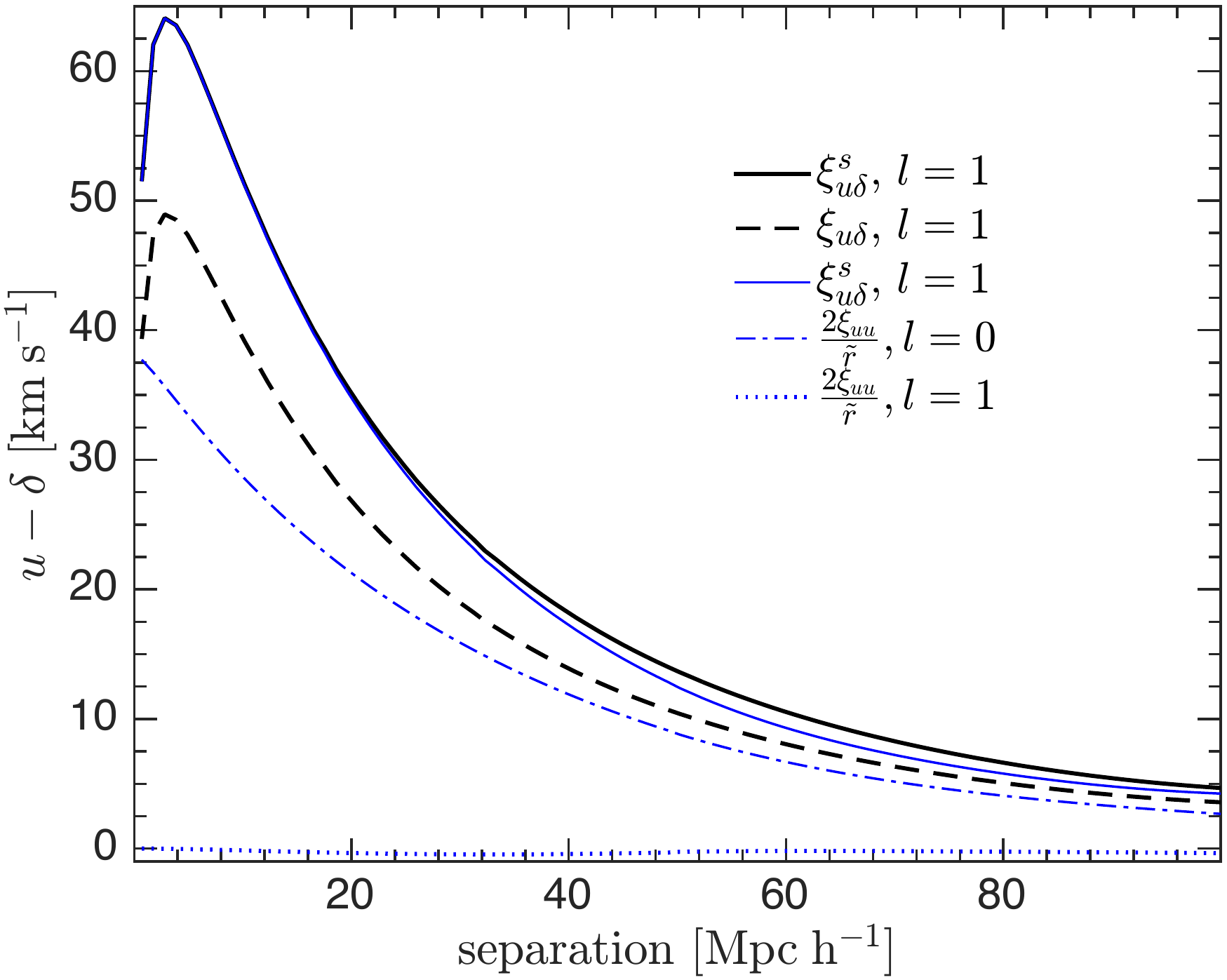}
\caption{Theoretical multipole moments of various terms in $u-\delta$ correlations, as indicated in the figure.
The black  solid and dashed curves are, respectively, the redshift and real space dipole ($l=1$) moment of the 
correlation in  the distant observer limit with $r=\infty$. 
The blue lines are for the case where the  $u$ is given  at a distance $ r=50\hmpc$ from the origin.}
\label{fig:theory}
\end{centering}
\end{figure}

In figure \ref{fig:theory} we plot a few of the relevant  moments computed from the relations above for the $\Lambda$CDM cosmology with the 
Planck parameters.
The multipoles $l=2$ and $3$ are much smaller than the others and we opt not to show them.
The $l=0$ moment of $\dsc{\xi}{uu}/\tr$  is substantial for $r=50\hmpc$. 
The  multipole $l=1$ of this term vanishes identically only in the DOL.  Nonetheless, it
 remains insignificant even away from this limit, as  seen from the blue dotted curve  at separations $\gsim r$.
Redshift space distortions  boost the correlations  with respect to real space due to large scale motions coherent with the density. This is reflected in the higher amplitude of $\la \dsc{\xi}{u\delta}^s\ra_{l=1}$ (solid black for DOL and solid blue for $r=50\hmpc)$) relative to the real space   $\la \dsc{\xi}{u\delta}\ra_{l=1}$ (dashed).
As we shall see, the difference between the DOL and the $r=50\hmpc$ results is insignificant compared to the errors associated with observations. The Kaiser rocket effect resulting from the flux limit nature of galaxy redshift surveys 
amounts to a term of the form $\dsc{\xi}{uu}/\tr$, which is negligible in  the $l=1$ moment of the correlation.

\begin{figure*}
\begin{centering}  
\includegraphics[angle=0,width=.8\textwidth]{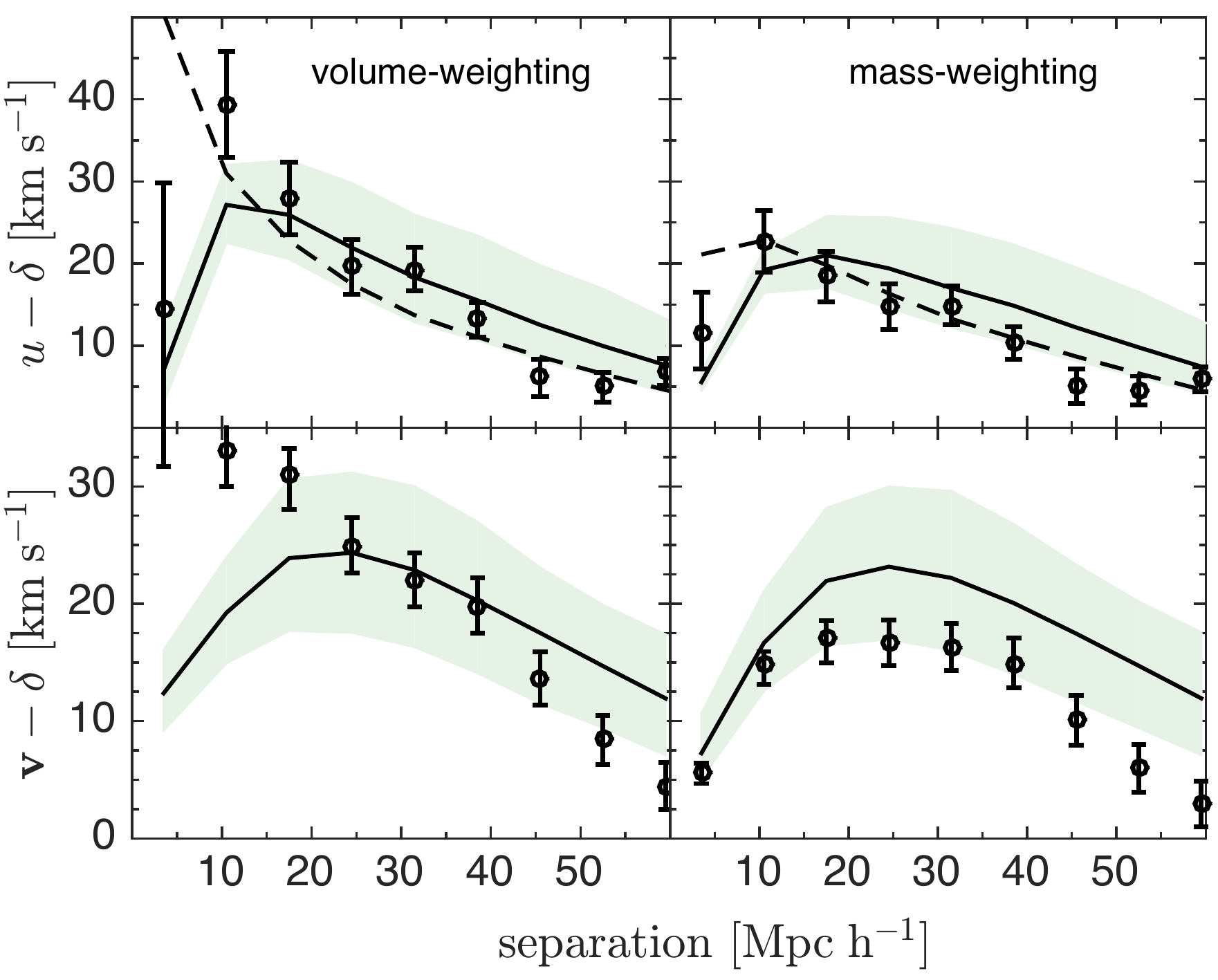}
\caption{\textit{Top:} Dipole moments, $\zeta$, of the correlations $\dsc{\xi}{u\delta}^s $ versus the separation, for  the  data (circles) and the mocks (solid and dashed in redshift and real space, respectively). The ($1\sigma$) error bars 
represent the scatter due random error in  \textit{cosmicflows-3} velocities. Colored areas mark $1\sigma$ cosmic variance and 
shot noise in 2MRS, as given from the mocks. Panels to the left and to the right show  $\zeta^V$ and $\zeta^M$,
respectively. \textit{Bottom:}  The same as \textit{Top}   but for 3D velocities reconstructed  using  Wiener filtering. }
\label{fig:datauV}
\end{centering}
\end{figure*}

\section{The observations}
\label{sec:obs}
\subsection{The distance indicator  catalog}
We use the grouped version of the \textit{cosmicflows-3} catalog \citep{cf3} consisting of groups of galaxies as well as  individual galaxies unassociated with groups. The total number of  objects in the catalog is 6560 with 
${\sim}6200$ having  redshifts $cz\ltsim 15000\kms$.
The catalog provides measurements of redshifts, $cz$,  and distance moduli, $
 DM=25+5\log_{10}(D_L/\mathrm{Mpc})
$
  where $ D_L$ is the luminosity distance.  To infer the radial peculiar velocities, $u$, we follow \cite{cf3}  who use a prescription \citep{ND95,ND11a,watkins} that avoids introducing spurious non-gaussian errors at large distances.

Uncertainties in the determination of $u$  increase linearly with  distance where  the error in groups is reduced 
by the square root of  the number of galaxies they respectively contain.
Galaxies at large distances are prone to suffer from  systematic biases and  therefore we choose to restrict our analysis to objects within a maximum redshift  of 
$10000\kms$, leaving us with ${\sim}4600$  objects.

\subsection{The redshift survey}
The distribution of galaxies in the flux limited  2MRS  catalog \citep{2mrs2012} will be correlated with peculiar velocities in \textit{cosmicflows-3}.
 The  2MRS   covers most of the sky down to Galactic latitudes of $b\le 5^\circ$ ($8^\circ$ in the direction of the Bulge) 
 and contains ${\sim}45000$ galaxies with spectroscopic redshifts 
and with apparent magnitudes   $K_s \le 11.75$. 
In order to account for the missing galaxies due to the flux limit  each galaxy is assigned a weight 
\begin{equation}
w=1/\Phi(r) \; ,
\end{equation}
where $\phi(r) $ is the selection function obtained from the $F/T$ estimator \citep{turner79,DH82}
taking into account evolution and k-correction \cite[figure 1 in ][]{branchini2012}. 
The galaxy number density contrast in a small volume, $\Delta V$,  is then estimated by 
\begin{equation}
\label{eq:deltadef}
\delta^\mathrm{g}=\frac{1}{\bar n} \sum_{\vr_\beta \in \Delta V} w_\beta -1 \; ,
\end{equation}
where $w_\beta=w(r_\beta) $ and the  sum is over all galaxies in $\Delta V$. Further,
\begin{equation}
\bar n=\frac{3}{4\pi R^3}\sum_{r_\beta <R} w_\beta 
\end{equation}
is a proxy to the mean number density of galaxies, computed from the distribution of galaxies out to a sufficiently large radius $R$. We take $R=120\hmpc$, but $\bar n $  is  robust for  $80\hmpc \ltsim R\ltsim 200\hmpc$, changing by less than a few percents. 

\subsection{Estimating correlations from the observations}
\label{sec:estimating}
Following the discussion in \S\ref{sec:theory}, we  consider the dipole ($l=1$) moment only. 
We bin the separation, $y$, into $N_b$ shells with radii  $y_b$ ($b=1\cdots N_b)$ of identical widths $\Delta y$.
For each galaxy, $\alpha$, at $\vr_\alpha$  in the \textit{cosmicflows-3} catalog we compute 
\begin{equation}
\label{eq:Qalpha}
 Q_\alpha=\sum_{y_{\alpha \beta}\approx y_b} w_\beta \mu_{\alpha\beta}
\end{equation}
where  $\beta$ indicates  2MRS galaxies at $\vr_\beta$,    $\vy_{\alpha\beta}=\vr_\beta-\vr_\alpha$ and
 $\mu_{\alpha\beta}=\vy_{\alpha\beta}\cdot \vr_\beta$ is  cosine the angle between $\vy_{\alpha\beta}$ and $\vr_\beta$. The approximate symbol in the expression above implies that the summation is over galaxies with $|y_{\alpha\beta}-y_b|<\Delta y$.
We consider two quantities,
 \begin{equation}
 \label{eq:zetaV}
\zeta^{V}(y_b)=\frac{1}{N^\mathrm{v}}\sum_\alpha \frac{Q_\alpha}{\bar nV_b} u_\alpha
\end{equation}
and 
\begin{equation}
 \label{eq:zetaM}
{\zeta^M}(y_b)=\frac{1}{N^\mathrm{v}}\sum_\alpha \frac{Q_\alpha}{\bar nV_b(1+\delta^\mathrm{g}_b)} u_\alpha
\end{equation}
where $N^\mathrm{v}$ is the number of objects in the velocity catalog, $\bar n V_b$ is the mean number of galaxies in the volume $V_b$ occupied by the corresponding shell, and 
$\delta_b$ is the mean density contrast inside  $V_b$.
Both of these quantities reflect  the dipole moment of $\dsc{\xi}{u\delta}^s$, but they have different physical meanings. While $\zeta^V$ correlates $u(\vr)$ with the dipole moment of the density in  spherical shell  centered on $\vr$, the function $\zeta^M$ correlates $u$ with the mean $\mu$ per unit mass in the shell, i.e. it gives the same weighting to 
shells of different densities, $\delta^\mathrm{g}_b$. The function $\zeta^V$ matches the basic   definition of the $l=1$ multipole of the $u-\delta$ correlation, while $\zeta^M$ significantly deviates from it 
in the non-linear regime  where $|\delta_b|\sim 1$. Nonetheless,  $\zeta^M$ should serve as  important test for the robustness of the results.

Let us explore these correlations in more detail.
In the continuous limit 
\begin{eqnarray}
Q_\alpha(y)&\rightarrow & \frac{\bar n V_b}{2} \int_{-1}^{+1}\left[1+\delta^\mathrm{g}(\vr_\alpha+\vy)\right] \mu \dd \mu\\
\nonumber &=&\frac{\bar n V_b}{2} \int_{-1}^{+1} \delta^\mathrm{g}(\vr_\alpha+\vy)\mu \dd \mu
\end{eqnarray}
where now $\mu= \hvr_\alpha \cdot \hvy$ and  $\delta^\mathrm{g}(\vr_\alpha +\vy)$ is the density contrast of the 2MRS galaxy distribution at the point $\vr_\alpha+\vy$ in the shell. Hence,
 \begin{equation}
 \label{eq:xivc}
\zeta^V(y)\rightarrow \frac{1}{2N^\mathrm{v}}\int _{V^\mathrm{v}}\dd^3 \vr n^\mathrm{v}(\vr)u(\vr)\int_{-1}^{+1}\delta^\mathrm{g}(\vr +\vy) \mu \dd \mu\; ,
\end{equation}
where $V^\mathrm{v}$ is the total volume encompassing the peculiar velocity  catalog and $n^\mathrm{v}(\vr) $
 is  the local number density 
of objects in the catalog,  expressed   as 
\begin{equation}
\label{eq:snu}
n^\mathrm{v}(\vr)= \bar n^\mathrm{v}  \left[1+\delta^\mathrm{v}(\vr)\right] S^\mathrm{v}(\vr)\; ,
\end{equation}
where $\bar n^\mathrm{v}$ is the underlying mean number density of objects and $\delta^\mathrm{v}$ is the underlying 
density contrast. Clearly $ \delta^\mathrm{v}$  and $\delta^\mathrm{g}$ are closely related since they represent 
the underlying distribution of galaxies, albeit of different populations. The function $S^\mathrm{v}(\vr) $ represents  various observational selections imposed on objects 
included in the velocity catalog. In contrast to flux limited redshift surveys, selection criteria on  distance measurements 
are diverse and in general cannot be represented in a well defined form. 
However, it is reasonable to assume that  none of these criteria depend  on $\delta^\mathrm{v}$. 
Hence, $S^\mathrm{v}$ is independent of $\delta^\mathrm{v}$ and  
\begin{equation}
N^\mathrm{v}=\bar n^\mathrm{v} \int\dd^3  \vr S(\vr) \; .
\end{equation}
Taking the ensemble average of the expression (\ref{eq:xivc}) we obtain
\begin{eqnarray}
\zeta^V&\rightarrow&\frac{\bar n^\mathrm{v}}{N^\mathrm{v}} \int \dd^3 \vr S(\vr) \zeta(r,y)\\
\nonumber 
&+& \frac{\bar n^\mathrm{v}}{2N^\mathrm{v}} \int \dd^3\vr S(\vr) \la \delta^\mathrm{v}(\vr) \delta^\mathrm{g}(\vr +\vy) u(\vr)\ra \mu \dd \mu\; .
\end{eqnarray}
 The second term   reflects a high order  correlation between the  density
and velocity fields. We do not focus on  this term although it is included in the 
correlations computed from the mocks used for comparison with data. This term is small on large scales.
In the first term
\begin{equation}
\zeta(r,y)\equiv \frac{1}{2}\int\dsc{\xi}{u\delta}^s(\vr,\vy) \mu \dd \mu \;.
\end{equation} 
Thus the first term is  an average of the  dipole moment of the underlying  linear correlation.
Since $\zeta$ is independent of the direction of $\vr$,  this term is expressed as 
\begin{equation}
\frac{4\pi \bar n^\mathrm{v}}{N^\mathrm{v}} \int \dd r r^2 \bar S(r) \zeta(r,y)\; ,
\end{equation}
where $\bar S$ is the angular average of $S(\vr)$, which is simply proportional to the radial distribution of objects in the  velocity catalog and it is easily modeled. Further,  in the distant observer limit $\zeta(r,y)\approx \zeta(y) $
and we have $\zeta^V \rightarrow \zeta$.
To linear order $\zeta^M=\zeta^V$ despite the different physical meanings they carry. On small scales where deviations 
from linear theory are seen, we expect a smaller amplitude for $\zeta^M$ as a result of weighting by  the density.

In principle we could assign each \textit{cosmicflows-3} galaxy in this sum a weight determined by the   observational  error\footnote{Observational errors include 
intrinsic scatter in the distance indicator as well as purely instrumental measurement errors},  $\sigma^\mathrm{v}$,  in $u_\alpha$.  Since the errors increase linearly with redshift, this procedure will 
heavily weigh nearer objects. The mean redshift of the \textit{cosmicflows-3} objects  (survey depth)
\begin{equation}
\overline{cz}={\sum_{cz_\alpha< cz_\mathrm{max}} w^\mathrm{v}_{\alpha} cz_\alpha}/{\sum w^\mathrm{v}_{\alpha}}
\end{equation}
is  ${\sim}1000\kms $ and $5000\kms$, respectively,  for $ w^\mathrm{v}=(\sigma^\mathrm{v})^{-2}$ and $w^\mathrm{v}=(\sigma^\mathrm{v})^{-1}$, both for 
 unrestricted $cz_\mathrm{max} $. Results in this paper will be given with $w=1$ including objects with  
 within $cz_\mathrm{max}=10000\kms$,  corresponding  $\overline{cz}=8000\kms$. This equal weighting of all galaxies within $cz_\mathrm{max}$ comes at  the expense 
 of increased statistical error on the estimated correlations. 
 Fortunately, as we shall see, the main source of uncertainty in the determination of the correlation is 
 cosmic variance rather than velocity observational errors. 
 The reason for imposing the redshift cut, $cz_\mathrm{max}$ is to avoid potential systematic errors in the data, which are more likely at larger distances.

We place  objects (galaxies and groups) in the peculiar velocity catalog at their comoving distance positions,
$r(z)$,
computed from observed redshifts, $cz$,  rather  than  
distances, $r_\mathrm{e}$,  estimated from the distance indicator measurements. 
Although the inferred peculiar velocities, $u$,  are the same, the placement of objects at $r_\mathrm{e}$
introduces severe spatial Malmquist biases  associated with large random  errors   \citep{Lynden-Bell1988,sw95}.
Non-linear and incoherent motions on small scales spoil the one-to-one mapping between distance and redshift and play the role of random errors when objects are placed at their redshift coordinates. Thus Malmquist  biases are present in redshift space as well. These 
small scale motions are  ${\sim}200\kms$ 
 while distance errors are 
$ {\sim} (0.15-0.2) cz $. Thus, working in redshift space, especially at large distances greatly mitigates the effects of 
Malmquist biases limiting  them to scales of   a few Mpc anyways.  We will be interested in the correlation signal on scales of 10s of Mpcs, implying that the effect of small scale  motions is entirely negligible. Note that the  2MRS galaxy distribution is smeared out along the line of sight due to these  motions. This prevents the  recovery of the underlying signal on a few Mpc scale even with perfect peculiar velocity data.

A peculiarity of our local cosmological neighborhood  is the remarkable coherence  of 
the flow of galaxies within a few Mpcs around us \citep[e.g.][]{CF2}. With respect to the CMB frame of reference, the bulk flow around us starts with ${\sim}630\kms$ for the  LG  and smoothly reaches ${\sim} 270 \kms$  in  a sphere of 
radius $100\hmpc$ centered on the observer \citep{ND11a}.   
Thus, the distribution of galaxies  defined by redshifts with respect to the CMB exhibits 
a large spurious local inhomogeneity. To avoid this behavior we work with redshifts transformed to the LG frame of reference. However,  
a coherent    galaxy  distribution is not the only reason for 
choosing this  frame of reference. 
The LG motion is generated by the cumulative gravitational of the large scale structure  out to 
$\gtsim 300\hmpc$ \citep{bilicki11,Nusser2014} and beyond. These distant (external) structures contribute also to the  velocities of  \textit{cosmicflows-3} 
objects with respect to the CMB, as indicated by the value of 
 bulk flow within $cz_\mathrm{max}/H0=100\hmpc$. The main contribution of  external structures is to  the dipole moments of the velocity field on shells around us. 
 However, the dipole component of the velocity field \textit{relative to the LG motion}  is entirely determined 
 by the mass distribution between the LG and the shell \citep{ND94}.  Therefore, working in the LG frame will 
 exclude  a velocity component which is uncorrelated with  the   density field surrounding 
 an object in the \textit{cosmicflows-3} catalog.

\section{Mock Catalogs}
\label{sec:mocks}
We use 30 mock catalogs of 2MRS extracted by \cite{KN16} from the 
  Dark-Sky public Simulations \citep{dss} of the $\Lambda$CDM cosmology with
 $\Omega = 0.295$ , baryonic density parameter $\Omega_b = 0.0468$, $\Lambda =
0.705$, $h = 0.688$, and  \textit{rms} mass density fluctuations in spheres of $8\hmpc$ of $\sigma_8 = 0.835$.
The parameters are very close to the Planck values  \citep[e.g. column 4 of  Table 3 in ][]{Planck2015}.
These are  dark matter only simulations, but baryonic feedback is too weak to play  any significant  role over the relevant  level of accuracy 
and scales 
considered \citep{Hellwing2016a}
The simulations are available with 5 different resolutions. 
The mocks are based on the simulation 
of $4096^3$
particles in a periodic box of $800 \hmpc$
 on the side, corresponding
to a particle mass of $ 6.1\times 10^8 h^{-1}
M_\odot$.  We consider halos
containing no less than 50 particles,  giving a minimal halo mass
of $3.05 \times  10^{10} h^{-1}M_\odot$. Mock  ``observers" are  selected to reside in  host
halos  of mass   $\gtsim 5 \times 10^{11} h^{-1}M_\odot $ corresponding to a Milky Way halo mass
with no neighboring  rich clusters ($\gtsim10^{15} h^{-1}M_\odot$) within a distance of 20 $\hmpc$. 
Further,  the bulk  velocity
within spheres of  radii 3, 5, and 7 $\hmpc$ centered on the observers do not  differ by more
than $50 \kms$. 

Halos  above a mass threshold $M_h$ are identified as mock galaxies. The mass threshold depends on the 
distance from the observer in each catalog in order to match the 
selection function of the  2MRS galaxies  \citep{branchini2012}. Each  catalog is trimmed at a maximum distance 
of $200\hmpc$ and contains $\sim 42000$  galaxies similar to the real  2MRS.
 
 Similarly to $\sigma_8$ for the underlying mass density field, we define $\sigma_8^\mathrm{g}$ as the \textit{rms }
 of  the fluctuations in the galaxy number density contrast in spheres of $8\hmpc$ in radius. We also define 
 $b=\sigma_8^\mathrm{g} /\sigma_8$ as the galaxy bias factor.
Direct calculation of $\sigma_8^\mathrm{g}$ in the mocks in real space yields 
$b^{\mathrm mock}\approx  0.96$, i.e. the mock galaxies are slightly anti-biased.
We have  directly  calculated  $\sigma_8^\mathrm{g}$  in the 2MRS catalog from randomly placed spheres. 
The result depends slightly on the volume chosen to contain the spheres. 
We obtain $0.98<\sigma_8^{\mathrm g}<1.07$ for spherical volumes of radii $80-140\hmpc$.
Thus, we  adopt $\sigma_8^{\mathrm g}=1.02\pm 0.05$ \citep[cf.][]{w09}. We are interested in $\sigma_8^\mathrm{g}$ computed from the 
galaxy distribution in real space.  Assuming the ratio of $\sigma_8^\mathrm{g}$ in redshift to real space is the same as in the mocks 
we derive $\sigma_8^\mathrm{g}=0.93\pm 0.05$  for 2MRS in real space
Adopting the Planck  $\sigma_8=0.831$ \citep{Planck2015} for $\sigma_8$ of underlying mass, yields the 2MRS bias factor of $b^\mathrm{2MRS}=1.12\pm 0.06$. 

Mock \textit{cosmicflows-3} galaxies are selected at random the 2MRS mocks 
with a distance dependent probability designed to 
produce a radial distribution of objects,  consistent with the  \textit{cosmicflows-3} catalog. The mean number of galaxies in the mocks within $100\hmpc$
is 4300, similar to the observations. We refrain from adding errors to the mock \textit{cosmicflows-3} objects since the observed errors
vary depending on the type of distance indicator used and whether the object is a group or a individual galaxy. 
The effect of random errors on  the correlations is done via a ``bootstrap"  procedure described in the next section.
\section{Results}
\label{sec:results}
\subsection{$u-\delta$ correlations}
The top panels in figure \ref{fig:datauV} show the correlations $\zeta$ computed from the 
data  and the mocks   as a function of separation in bins of $7\hmpc$ in width.
The data correlations are  represented by the circles and the attached  error bars  correspond to the expected $rms$ scatter due to  random observational errors on  peculiar velocities of individual  \textit{cosmicflows-3} objects. 
In order to obtain the error bars, we have perturbed the peculiar velocity of every object in  \textit{cosmicflows-3} 
by  a random variable drawn from a normal distribution with $rms$ equal to the $1\sigma$ velocity error on that object. We have generated a 100 of these bootstrap realizations of the data and computed the 
corresponding correlation functions. The error bars in the figure are  $rms$  values of scatter 
of these correlations from the mean.
Thus these  error bars  do not include cosmic variance and 
Poisson noise  (shot noise) in the 2MRS discrete galaxy distribution. 
The modeling of these two effects is done via the mock catalogs. The mean correlation functions  from  the 30 mocks are shown as the solid and dashed curves for 
redshift and real space, respectively.
The light colored shaded areas are the 
$rms $ scatter around the solid curve computed from the 30 mocks in redshift space. Thus the shaded 
areas include cosmic variance as well as shot noise (but without observational errors on $u$).

At scales $\ltsim 15\hmpc$,  the volume-weighted correlation $\zeta^V$  (panel to the left)  is larger than $\zeta^M$ (panel to the right)  in the  mass weighting scheme. However, on larger scales, the match between the two 
functions is good.
The data points for  both $\zeta^V$ and  $\zeta^M$
match  well the corresponding correlations in the mocks. 
Comparison between the mock correlations  in real and redshift space, indicates that the decline of the correlations below separations $\ltsim 15 \hmpc$ is mostly caused by the smearing out of 
structure in redshift space as a result of nonlinear and incoherent motions on small scales.  On larger scales, the mock  correlations are stronger in redshift space than in real space, as expected from the enhanced apparent  large scale clustering in redshift space due to 
coherent motions.
 On scales $\gtsim 20\hmpc$,  there is a good  agreement between the data and the mock on one hand and the theoretical correlations in figure \ref{fig:theory} on the other hand. 
 This is reasonable since linear theory is expected to be valid on these scales and the 
 galaxy bias factor in the mocks and the 2MRS  is $b\sim 1$.

The total error covariance  in the estimation of the correlation functions 
is  obtained by   adding the covariance matrices  from the mocks 
 and the  bootstrap realizations. 
Using the  covariance matrix, $C_i$, for the 9 measurement points in each of the top panel of figure (\ref{fig:datauV})
we compute 
\begin{equation}
\chi^2=\zeta_i C^{-1}_{ij} \zeta_j\; ,
\end{equation}
where $\zeta_i$ is the data correlation. This $\chi^2$ determines whether 
the \textit{null} hypothesis of  \textit{zero} underlying correlation can be ruled out or not within a certain significance level.
We find  $\chi^2=38.67  $  and $34.85$ for $\zeta^V$ and $\zeta^M$, respectively. 
For  9 degrees of freedom (dof), the $\chi^2$ distribution function yields a rejection of better than $1.32\times 10^{-5}$ 
and $6.32\times 10^{-5}$ for $\zeta^V$ and $\zeta^M$, respectively.
Expressed in terms of a normal distribution of a single variable, these 
rejection levels, respectively,  correspond to  $4.35 \sigma$ and $4\sigma$.
The agreement with the predictions of the $\Lambda$CDM model as  given by the mocks can similarly be assessed by  the same form  $\chi^2$ computed  with   $\zeta-\zeta_{\Lambda \mathrm{CDM}}$ instead of $\zeta$. Here 
$\zeta_{\Lambda \mathrm{CDM}}$ is  the mean of the mock  correlations, represented by the 
solid curves in the figure. We obtain $\chi^2=14.97$ and 11.13 for volume and mass weighting, respectively. 
These values correspond to $1.68\sigma$ and $1.11\sigma$ levels, making the $\Lambda$CDM entirely consistent with the correlation measurements.
 
\subsection{Constraints on the growth rate and clustering amplitude}

Predictions of  $u-\delta$ correlations from a cosmological model depend on all  cosmological parameters. 
We have seen that the correlation is consistent with the $\Lambda$CDM model with the Planck  cosmological parameters \citep{Planck2015}. Here we aim at deriving constraints on the normalization of the density fluctuations, $\sigma_8$ and the growth rate, $f$, assuming the Planck values for all other cosmological parameters.

\begin{figure}
\begin{centering}  
\includegraphics[angle=0,width=.5\textwidth]{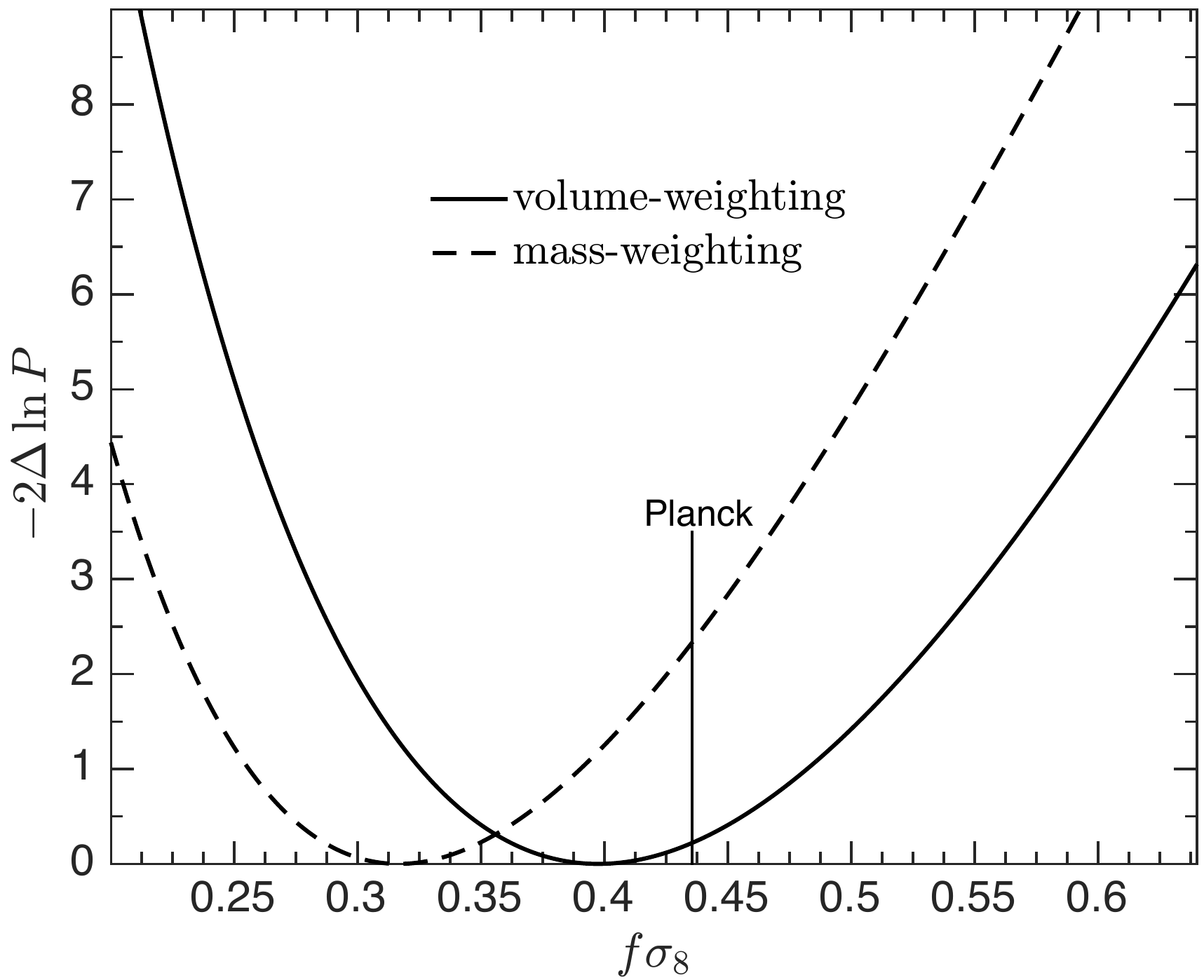}
\caption{Curves of $-2\Delta \ln P $ (equation \ref{eq:chif}) versus $f\sigma_8$ derived from the observed $\zeta^V$ (solid) and $\zeta^M$ (dashed).}
\label{fig:fsig}
\end{centering}
\end{figure}

The linear theory relation (\ref{eq:linr}) between the velocity and the mass density, implies the scaling  $u \propto f \sigma_8 $ of the amplitude of $u$ on 
the combination $f\sigma_8$. From this scaling and the expression (\ref{eq:xisud}) it is easy to derive the dependence of  $\dsc{\xi}{u\delta}^s$ on these parameters.  The correlation  $\dsc{\xi}{u\delta}^s$ involves the galaxy number density contrast with 
$\sigma_8^\mathrm{g}=b\sigma_8$, hence,  in (\ref{eq:xisud}) 
$\dsc{\xi}{u\delta}\propto b f \sigma_8^2 =f\sigma_8 \sigma_8^\mathrm{g}$  and  $\dsc{\xi}{uw}\propto (f\sigma_8)^2$. Therefore, 
\begin{equation}
\label{eq:ximodel}
\dsc{\xi}{u\delta}^s=f \sigma_8 \sigma_8^\mathrm{g}\dsc{\xi}{u\delta}^1 +(f\sigma_8)^2\dsc{\xi}{uw}^1
\end{equation}
 where the  superscript ``1" indicates  correlations 
computed for $f\sigma_8=1$ and  we have omitted  the $\dsc{\xi}{uu}$ term which does does not contribute to the dipole moment.
The correlations $\dsc{\xi}{u\delta}^1$ and $ \dsc{\xi}{uw}^1$ could easily be estimated either from the theoretical expressions 
in \S\ref{sec:theory} or the mock catalogs. Here we work with the mocks since they provide a more realistic 
description of the data. We extract $f\sigma_8$ by minimizing the function
\begin{eqnarray}
\label{eq:chif}
-2\ln P&=&(\zeta_i-\zeta^{\mathrm{m}}_i) (C^{\mathrm{m}})^{-1}_{ij} (\zeta_j-\zeta^{\mathrm{m}}_j)\\
\nonumber &+& \ln \left[\mathrm{det} \left(C^{\mathrm{m}}\right)\right]
\end{eqnarray}
where  
$\zeta^{\mathrm{m}}(f\sigma_8) $  is the model prediction   obtained according to (\ref{eq:ximodel}) using  $\dsc{\xi}{u\delta}^1$ and $ \dsc{\xi}{uw}^1$ of the mocks. 
The covariance matrix $C^\mathrm{m}$ depends on $f\sigma_8$ through the cosmic variance term which 
is also extracted from the mocks assuming it has the same $f\sigma$ scaling as the correlations.
Since we are using linear theory to model the dependence on  $f\sigma_8$, we limit the calculation of $\chi^2$ in (\ref{eq:chif})  to large scales,  including only the 7 points with separations larger than 15$\hmpc$.
 Further, we take  $\sigma_8^\mathrm{g} \approx 0.93\pm 0.05$ (cf. \S\ref{sec:mocks})   in (\ref{eq:ximodel}).
 Figure \ref{fig:fsig} plots curves of $-2\Delta \ln P=-2\ln P +2 \ln [\mathrm{max}(P)]$ for the volume and mass weighting schemes, as indicated in the  figure.
Minima  of $-2\Delta\ln P$  are rendered at the  best fit values of $f\sigma_8$ and 
and  $-2\Delta\ln P=1$ mark the corresponding $1\sigma$ errors.
 We find
 \begin{equation}
 f\sigma_8=\begin{cases}
 0.40\pm 0.08 & \text{for } \zeta^V\\
 &\\
 0.32\pm 0.07  & \text{for } \zeta^M
 \end{cases}
 \end{equation}
 which are both consistent with the Planck value $f\sigma_8=0.43$.
 These estimates can be expressed in terms of $\beta\equiv f/b=f\sigma_8/\sigma_8^{\mathrm g}$.  For  $\sigma_8^{\mathrm g}= 0.93\pm 0.05$ we obtain,
\begin{equation}
 \beta=\begin{cases}
 0.43\pm 0.09 & \text{for } \zeta^V\\
 &\\
 0.34\pm 0.08 & \text{for } \zeta^M
 \end{cases}
 \end{equation}
The parameters obtained from the two curves are consistent at the $\sim 1\sigma $ level. It is tempting to derive parameters based on the joint 
measurements of $\zeta^V$ and $\zeta^M$. Unfortunately, the number of mocks we have do not allow 
a robust estimation of the strong  covariance between $\zeta^V$ and $\zeta^M$. Attempts to do that with the current mocks, lead to 
a singular covariance matrix.

\subsection{$\vv-\delta$ correlations: preliminary assessment }
\label{sec:vvdelta}
\begin{figure}
\vspace{2cm}
\begin{centering}  
\includegraphics[angle=0,width=0.5\textwidth]{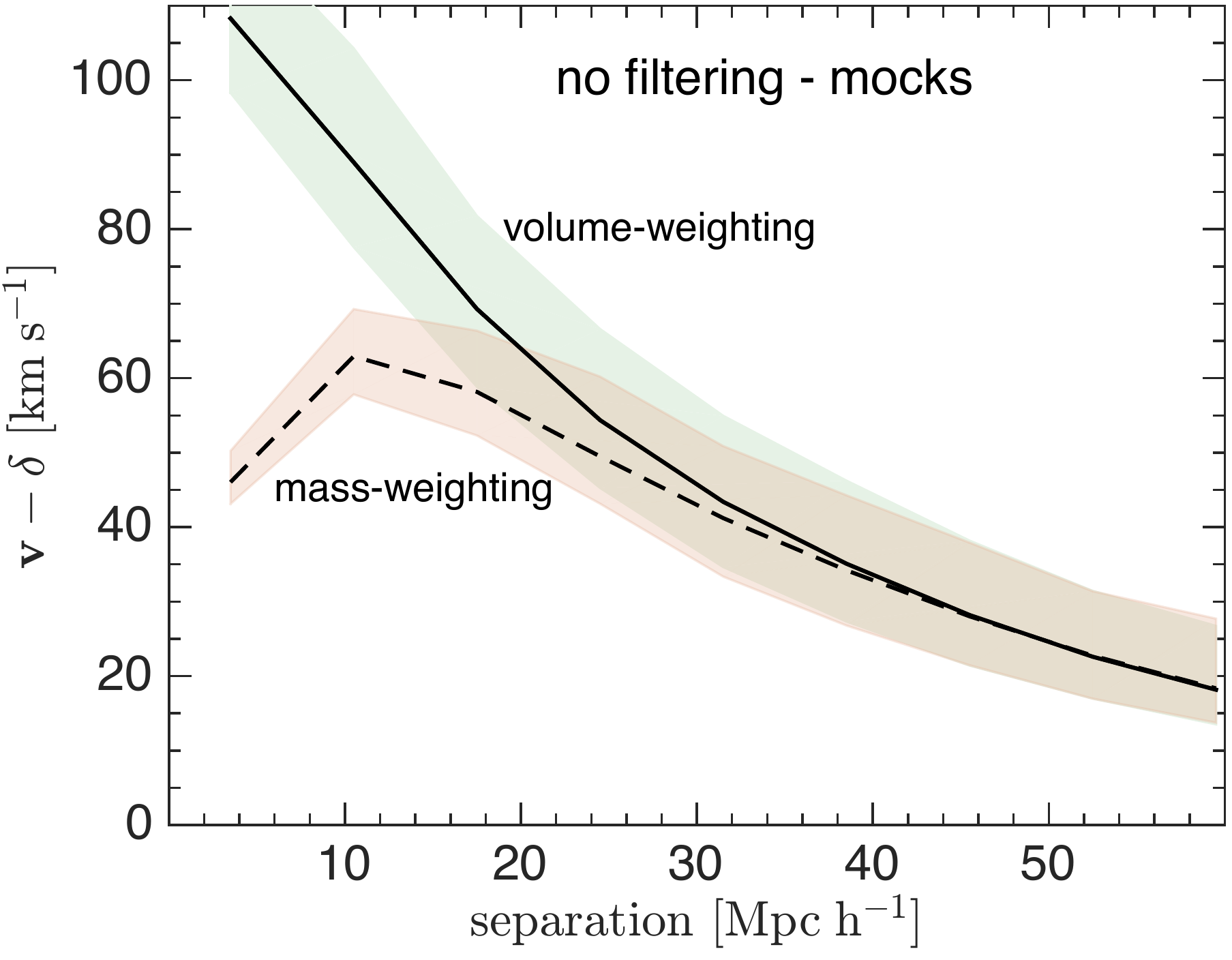}
\caption{Dipole ($l=1$) moments  of $\vv-\delta$ correlations derived from unfiltered peculiar velocities in the mocks. }
\label{fig:xi3d}
\end{centering}
\end{figure}

The 3D velocity is better correlated with  the surrounding density fluctuations. This is mathematically evident  by the expressions for $\dsc{\vxi}{\vv \delta}$ and $\dsc{\xi}{u \delta}$ in \S\ref{sec:theory}.
Figure \ref{fig:xi3d} illustrates this point. The solid and dashed curves are the dipole moments of the 
$\vv-\delta $ correlation from  mocks, respectively, for volume and mass weighting schemes. The shaded areas are the corresponding cosmic variance and shot noise scatter.
The curves are obtained in  a similar way to the calculation of  $\zeta$ for $u-\delta$, as described in \S\ref{sec:estimating}, with 
two differences:   $\mu_\beta$  is replaced with $\hvy\cdot \vv_\alpha/v_\alpha$ in (\ref{eq:Qalpha}) and $u_\alpha$ is replaced with $v_\alpha$ in both (\ref{eq:zetaV}) and (\ref{eq:zetaM}). 
Comparison with the $u-\delta$ correlations in the top panels in figure \ref{fig:datauV}, clearly reveals  a stronger 
$\vv-\delta$ signal  for a similar width of the shaded areas. 

The data provides the radial velocities and
we need a method for inferring the underlying 3D velocity from sparse and noisy-data.
Under the assumption of potential flow \citep{POTENT}, the 3D velocity field can be reconstructed from the observed radial peculiar motions. Observational errors and incomplete spatial coverage in velocity catalogs limit the inference of the 3D flow to  large scales only. 
We employ the methodology of Wiener filtering \citep{zh95} to reconstruct the 3D velocities.
Given a set of measurements $d_i$  at  points $\vr_i$ $(i=1\cdots N)$ the 
Wiener filtered (WF)  field of a desired signal, $F^\mathrm{WF}(\vr)$  at any point $\vr$ is 
\begin{equation}
F^\mathrm{WF}(\vr)=\Xi_{r \alpha'} {\mathcal{D}}^{-1}_{\alpha\alpha'} d_\alpha
\end{equation}
where $\Xi$  and $\mathcal{D }$ are the ``data-data"  and ``data-signal" correlation matrices. For gaussian errors and 
a gaussian underlying signal, the field $F^\mathrm{WF}$ renders a maximum in the 
probability $P(F|data)$ 
for $F$ given the data.  In our case, $F$ is the 3D velocity field, $\vv$, and $d$ represents the observed radial velocities $u$. Therefore, $\Xi=\dsc{\xi}{u\vv}$ and $\mathcal{D}=\dsc{\xi}{uu}+\mathcal{N}$ where $\mathcal{N}$ is the (diagonal) error 
matrix of $u$ in \textit{cosmicflows-3}. The calculation of these matrices is done using the expressions derived in \S\ref{sec:theory}, modified for velocities in the LG frame. Thus instead of $\dsc{\xi}{uu}=\la u(\vr) u(\tvr)\ra$ we 
consider $\la (u(\vr)-\vv(0)\cdot \hvr)(u(\tvr) -\vv(0)\cdot \htvr) \ra $ and similarly for $\dsc{\xi}{u\vv}$.
Im addition to $\vv$, we also apply the  WF methodology to derive the most 
probable real space density contrast field, $\delta$, given the 
\textit{cosmicflows-3} data. This is done by taking $\Xi=\dsc{\xi}{u\delta}$ where the $u-\delta$  correlation is given in (\ref{eq:thetau}).

\begin{figure}
\begin{centering}  
\vskip 1cm
\includegraphics[angle=0,width=0.45\textwidth]{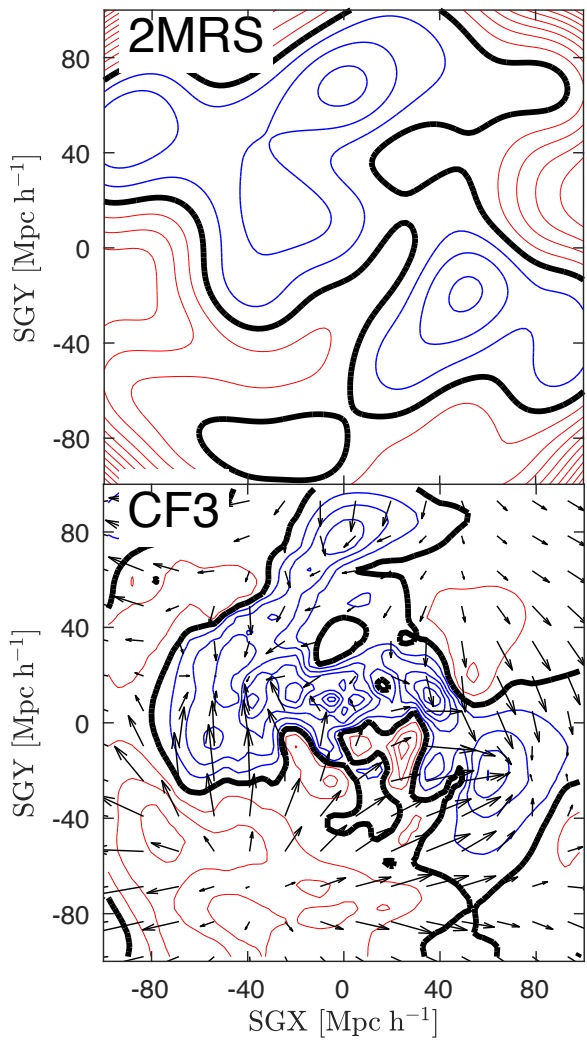}
\caption{\textit{Top:}  Contour map of the natural logarithm of the 2MRS density field smoothed with a Gaussian window of $10\hmpc$ in width. \textit{Bottom:} The velocity field $\vv^\mathrm{WF}$ overlaid on 
maps of  $\delta^\mathrm{WF}=-\vnabla\cdot \vv^\mathrm{WF}/f$,  obtained by Wiener filtering of   \textit{cosmicflows-3}. 
Arrow lengths  are proportional to the amplitude of the velocity where  the 
arrow nearest  to the bottom-left corner  corresponds to $360\kms$.
Both panels show fields in the supergalactic plane and  the contour spacing  is $0.25$.  Black, blue and red contours correspond to zero, 
positive  and negative  density contrast, respectively.}
\label{fig:WF}
\vskip 1cm
\end{centering}
\end{figure}

\begin{figure}
\vspace{2cm}
\begin{centering}  
\includegraphics[angle=0,width=0.5\textwidth]{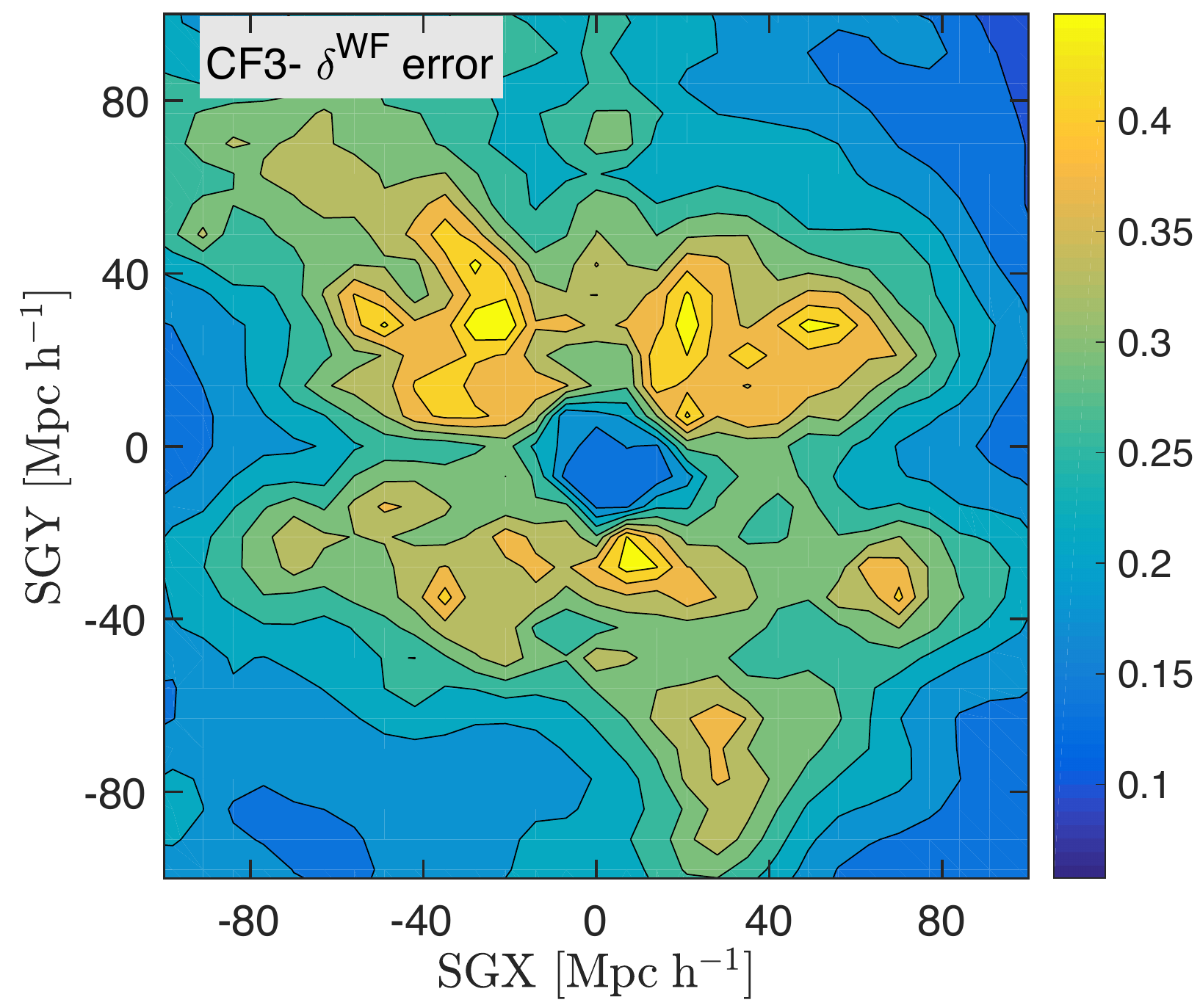}
\caption{Map of the expected $1\sigma$ error in $\delta^\mathrm{WF}$.}
\label{fig:WFerr}
\end{centering}
\end{figure}

\begin{figure}
\vspace{2cm}
\begin{centering}  
\includegraphics[angle=0,width=0.5\textwidth]{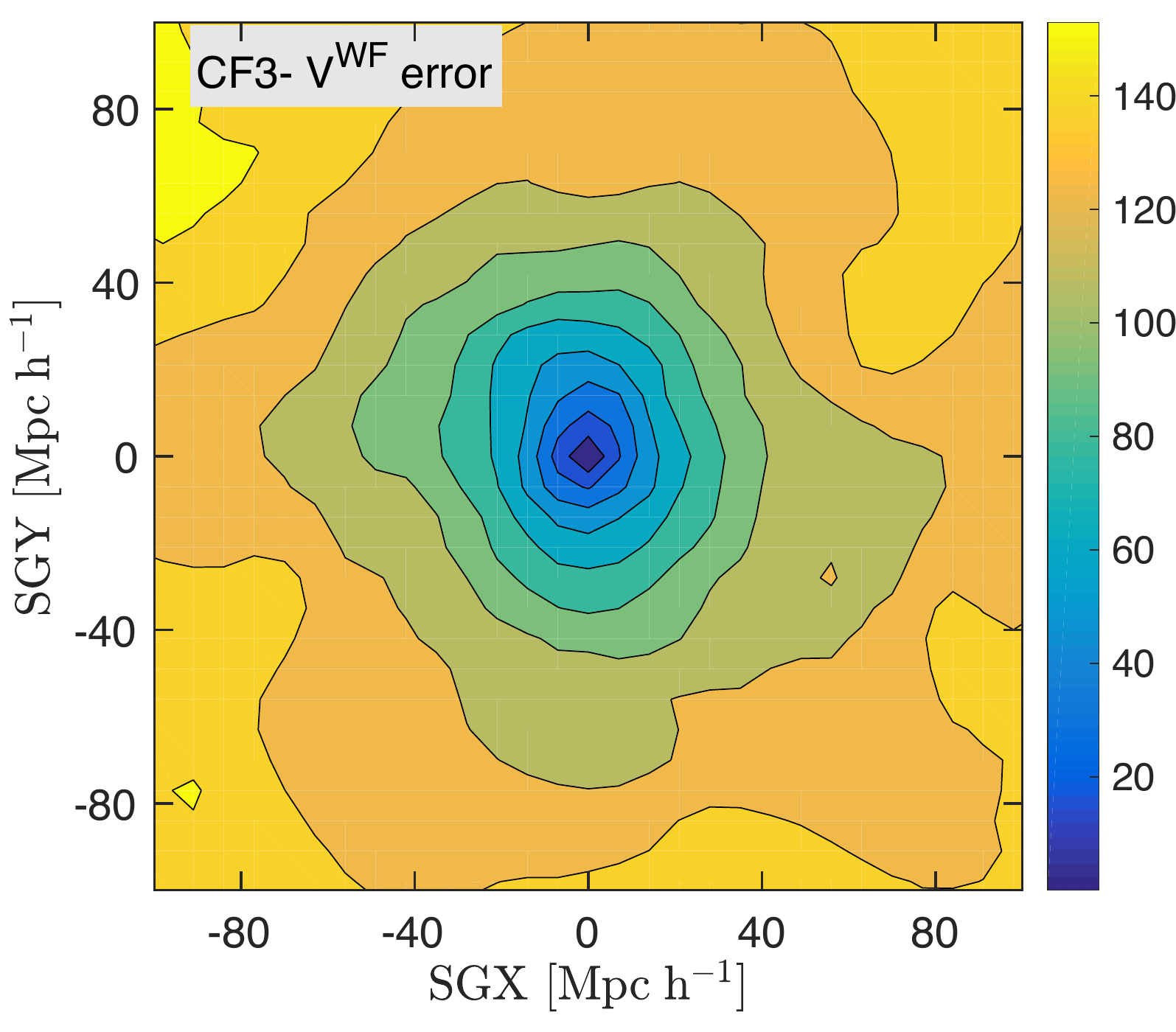}
\caption{Map of the expected $1\sigma$ error (in $\kms$)  in $\vv^\mathrm{WF}$.}
\label{fig:WFverr}
\end{centering}
\end{figure}

The WF  3D velocity, $\vv^\mathrm{WF}$ and the corresponding  density contrast, $\delta^\mathrm{WF}=-\vnabla\cdot \vv/f$ on a uniform grid  in the supergalactic plane (SGP)
are shown in the bottom panel figure \ref{fig:WF}. The mean velocity in the SGP has been subtracted in order to 
emphasize the flow pattern in the plane.
A comparison  the  WF reconstruction  of  \cite{Hoffman2015}  for  the \textit{cosmicflows-2} \citep{CF2} catalog 
reveals a good general match. The Perseus-Pisces is apparent as an extended over-density in the bottom-right corner, Coma at the top, a horizontally elongated over-density from $SGX=0 $ to $SGX=-80\hmpc$ at $SGY\approx 0$ and a large void in the bottom left corner.
 The top panel shows contours of $\ln(1+\delta)$ 
obtained by 
smoothing the 2MRS galaxy distribution. There is a reasonable visual  match between the 
$\delta^\mathrm{WF}$ and the 2MRS density. While a constant smoothing is applied to the 2MRS, the 
WF filtering is a strong  function of the errors and sparseness of the data. 
The agreement is satisfactory in light of the error map in figures \ref{fig:WFerr}\&\ref{fig:WFverr}. It is interesting that the errors in velocity increase with distance while decrease in the density. 

Using $\vv^\mathrm{WF}$, we compute the dipole moment of the 
velocity density correlation.
Volume and mass weighted dipole moments of $\vv^\mathrm{WF}-\delta$ are shown in the bottom panels of figure \ref{fig:datauV}.  The amplitude of the correlations,  both in the data and the mocks, is substantially lower than the amplitude of the unfiltered $\vv-\delta$ correlations of
in figure \ref{fig:xi3d}. 
 The shaded areas representing shot-noise and cosmic variances corresponding to  the $\vv-\delta$ correlation and shown 
 in the bottom panels of Figure \ref{fig:datauV} are broader than in the top panels for the $u-\delta$. This may seem surprising since a stronger 
 correlation is expected for the 3D velocity. The reason for that is that the derivation of the 3D velocity  is associated with 
 an interpolation of the radial velocity data to all spaces. This effectively introduces additional source of error, as also
 seen in  figures \ref{fig:WFerr}\&\ref{fig:WFverr}.
 
At separations $\gtsim 15 \hmpc$, the  data and the  mock  correlations are in very good agreement for 
the volume-weighted correlation (left panel). 
The overall data correlation  is below the mock curve in the mass weighting scheme (right panel).
 However, there is an important caveat in this comparison between data and mocks. 
 The WF introduces a specific form of smoothing which depends on the specific spatial  distribution of data points. 
To achieve an equivalent smoothing in the mocks, the \textit{cosmicflows-3} data points are  padded  in each mock and assigned the radial peculiar velocities of the nearest respective mock galaxies. Gaussian random scatter  is   then added according to the  error estimate in the \textit{cosmicflows-3}. 
This padding of data points in the mocks is done only for the purpose of $\vv-\delta $ correlation. It ensures 
identical WF window for  mocks as in data but misses 
the  correspondence between the selection of galaxies in the \textit{cosmicflows-3} and the underlying mass density. More galaxies should be selected in regions of higher density  where the flow is convergent, and vice versa. 
Thus the comparison presented here should only serve as a ``proof of concept"  motivating a more thorough analysis. 
A potentially fruitful path to proceed is to augment the $\vv^\mathrm{WF}$ with the missing small scale power 
according to \cite{HR91}.  For  Gaussian fields, this will yield the most likely unfiltered estimates of the  3D velocities, given the 
radial peculiar velocities. 


\section{summary and discussion}
\label{sec:conc}
Distance measurements  are very challenging and the  \textit{cosmicflows-3} catalog is under constant ``quality control"
by its observers. Ongoing improvements in the catalog are expected to reduce systematics, especially at large distances. 
Following private communications with the observers, we opted to limit the analysis to objects within
$cz_\mathrm{max}=10000\kms$. An attempt to compute the correlation including objects  at $cz>cz_\mathrm{max} $ failed to reach converging results.
As seen in figure \ref{fig:datauV}, it is cosmic variance rather than random observational errors that constitute the main source of random uncertainty in   $u-\delta$.   Suppression of  cosmic variance requires a significant  increase in  the 
volume probed by velocity catalogs. 
The next  \textit{cosmicflows-4} is expected
to  have a better control of systematics, allowing us to 
use  data at larger distances with greater confidence. Another important likely development is the likely  HI 
SKA precursor telescopes \citep[e.g.][]{Blyth2015}, which  could potentially provide  Tully-Fisher distances for tens of thousands  of galaxies within $300\hmpc$.

It is encouraging that the constraints derived here are consistent with other estimates based on 2MASS galaxies
\citep[e.g.][]{Pike05,Erdogdu06,Lavaux08,bilicki11,branchini2012}.
In particular, the detailed analysis of 
\cite{Davis2011} have demonstrated an excellent alignment  between the velocity field from the SFI++ catalog  of peculiar velocities and the gravitational force field obtained from the  $K_s=11.25$  2MRS of 
${\sim}2300$ galaxies. 
The alignment  is   best  for $\beta=0.33\pm 0.045$ which is in excellent agreement with  both 
$\beta=0.34\pm 0.08$ (mass weighting)  and  $\beta=0.43\pm 0.09$ (volume weighting).
This is perhaps not surprising since  \textit{cosmicflows-3}  relies heavily on the SFI++ and  the 
 $K_s=11.75$ 2MRS  is selected from the same  galaxy population as the $K_s=11.25$.
 Still  the agreement is impressive given the different methodologies.

Velocity-density  correlations play a role in the modeling  of redshift distortions in the density
auto correlation function as  estimated from redshift surveys \citep{Koda2014,okumura14,sugiyamaspergel}. Thus the analysis presented here could be useful since 
it provides direct observational constraints on these correlations. 

Velocity-velocity ($u-u$) correlations can be computed directly from peculiar velocity catalogs. Galaxy motions 
are  unbiased tracers of the underlying velocity field. Thus, in principle, $u-u$ correlations are independent of galaxy bias, unlike $u-\delta$ correlations. 
However, in practice $u-u$ correlations are  affected by the
distribution of objects  in the velocity catalog. Further, they
are prone to larger cosmic variance uncertainties \citep{Hellwing2017}. Therefore, $u-\delta$ correlations may prove more robust and useful.

 Peculiar velocities induce spatial modulations of the observed galaxy  luminosity function (when redshifts are used to transform fluxes to luminosities). Thus  spatial luminosity variations 
could serve as a probe of peculiar motions \citep{TYS,Nusser2011a,Feix2015,Feix2016}.
Thus useful information could be extracted  by correlating spatial deviations from the global luminosity function on one hand and 
the surrounding galaxy distribution on the other. This particular application of the luminosity modulation method
could be susceptible to environmental dependences of the luminosity function. Nevertheless, the application should at least  yield upper limits  on the amplitude of the correlations.
 
 While this paper is in the production stages, \cite{Adams2017} presented an analysis of the density-velocity covariance to improve cosmological constraints using 6dF data \citep{Jones2004}.

\section*{acknowledgments}
The author thanks Enzo Branchini, Marc Davis, Maciek Bilicki and Wojtek Hellwing for useful comments.
 This research was
supported by the I-CORE Program of the Planning and Budgeting
Committee, THE ISRAEL SCIENCE FOUNDATION (grants
No. 1829/12 and No. 203/09) and the Asher Space Research Institute.
 
 \bibliography{/Users/adi/Documents/Bibtex/library.bib}

\end{document}